\documentclass[aps, prl, reprint, showpacs, twocolumn, 10pt, superscriptaddress]{revtex4-2}

\usepackage{graphicx, amsmath, dsfont, dcolumn, bm, times, color, braket, amssymb, multirow}

\begin{document}
	
	\bibliographystyle{apsrev4-2}

	\title{Direct detection of odd-frequency superconductivity via time- and angle-resolved photoelectron fluctuation spectroscopy}
	
	\author{Viktoriia Kornich}
	\affiliation{Institut f\"ur Theoretische Physik and Astrophysik, Universit\"at W\"urzburg, 97074 W\"urzburg, Germany }
	\author{Frank Schlawin}
	\affiliation{Max Planck Institute for the Structure and Dynamics of Matter, Center for Free Electron Laser Science (CFEL), Luruper Chaussee 149, 22761 Hamburg, Germany}
	\affiliation{The Hamburg Centre for Ultrafast Imaging, Luruper Chaussee 149, 22761 Hamburg, Germany}
	\author{Michael A. Sentef}
	\affiliation{Max Planck Institute for the Structure and Dynamics of Matter, Center for Free Electron Laser Science (CFEL), Luruper Chaussee 149, 22761 Hamburg, Germany}
	\author{Bj\"orn Trauzettel}
	\affiliation{Institut f\"ur Theoretische Physik and Astrophysik, Universit\"at W\"urzburg, 97074 W\"urzburg, Germany }

	\date{\today}
	
	\begin{abstract}
		We propose a measurement scheme to directly detect odd-frequency superconductivity via time- and angle-resolved photoelectron fluctuation spectroscopy. The scheme includes two consecutive, non-overlapping probe pulses applied to a superconducting sample. The photoemitted electrons are collected in a momentum-resolved fashion. Correlations between signals with opposite momenta are analyzed. Remarkably, these correlations are directly proportional to the absolute square of the time-ordered anomalous Green's function of the superconductor. This setup allows for the direct detection of the ``hidden order parameter'' of odd-frequency pairing. We illustrate this general scheme by analyzing the signal for the prototypical case of a two-band superconductor.
	\end{abstract}

	\maketitle
	
	\let\oldvec\vec
	\renewcommand{\vec}[1]{\ensuremath{\boldsymbol{#1}}}
	
	{\it Introduction.--} Odd-frequency superconductivity is a genuinely dynamic state of matter. It is based on a pairing mechanism, in which the two electrons that form a Cooper pair in the superconducting condensate have to correlate with each other at unequal times. This is in contrast to the even-frequency case, when the superconducting pairing can be non-vanishing at coinciding times. The most well-known example of even-frequency superconductors are conventional superconductors with s-wave, spin singlet pairing. Odd-frequency  pairing was originally proposed by Berezinskii in the context of Helium-3 as a novel type of spin triplet pairing \cite{Ber1974}. Subsequently, it was realized by Balatsky and Abrahams that odd-frequency pairing could also exist in a spin singlet version \cite{Bal1992}, accompanied by other works that transferred the odd-frequency pairing concept to solid-state platforms \cite{kirkpatrick:prl1991,emery:prb1992,coleman:prl93,coleman:jpcm1997,fuseya:jpsj03}. In the meantime, a number of review articles have emerged that describe odd-frequency superconductivity from different perspectives \cite{Ber2005,Tan2012,Lin2019,Cay2020}. Moreover, certain experiments have been interpreted as indirect evidence of odd-frequency pairing.  Among the well-known examples are the measurement of a long-range supercurrent in Josephson junctions based on Nb, containing the strong ferromagnetic material Co \cite{Kha2010}, the observation of the polar Kerr effect in the heavy-fermion superconductor UPt$_3$ \cite{schemm:science14} (in combination with the theoretical analysis done in Refs.~\cite{wang:prb17,triola:prb18}), and the measurement of an intrinsic paramagnetic Meissner effect in Au/Ho/Nb trilayer systems \cite{DiB2015}. Additionally, it has been argued that signatures of odd-frequency pairing are visible in scanning tunneling spectra taken on top of magnetic impurities immersed in a Pb/Si(111) monolayer \cite{Per2020}. Recent theoretical proposals to detect odd-frequency pairing are, for instance, based on measurements of a supercurrent running from a Majorana STM to a superconducting substrate \cite{Kas2017}, transport properties through a quantum spin Hall system in proximity to an $s$-wave superconductor \cite{Fle2018}, the Josephson current on the surface of Weyl nodal loop semimetals \cite{Dut2019}, or the Josephson junction current noise \cite{Sou2020}.
	
	All these examples have in common (with the possible exception of Ref.~\cite{Sou2020}) that they can only be regarded as an indirect evidence for odd-frequency pairing for the following reason. The defining property of odd-frequency pairing is the mathematical (anti-)symmetry of the time-ordered, anomalous Green's function $F_{\sigma,\sigma'}^{{\bm p},{\bm k}}(t_1,t_2) \equiv \langle Tc_{{\bm p},\sigma}(t_1)c_{{\bm k},\sigma'}(t_2)\rangle$ (for fermionic annihilation operators with momentum $\bm p$, $\bm k$ and spin $\sigma$, $\sigma'$, respectively) under the exchange of the two time arguments $t_1$ and $t_2$. If $F_{\sigma,\sigma'}^{{\bm p},{\bm k}}(t_1,t_2)$ has an odd frequency component, which we name $F_{o,\sigma,\sigma'}^{{\bm p},{\bm k}}(t_1,t_2)$, then the relation $F_{o,\sigma,\sigma'}^{{\bm p},{\bm k}}(t_1,t_2) = - F_{o,\sigma,\sigma'}^{{\bm p},{\bm k}}(t_2,t_1)$ holds. In contrast, for the even frequency component $F_{e,\sigma,\sigma'}^{{\bm p},{\bm k}}(t_1,t_2)$, the relation $F_{e,\sigma,\sigma'}^{{\bm p},{\bm k}}(t_1,t_2) = F_{e,\sigma,\sigma'}^{{\bm p},{\bm k}}(t_2,t_1)$ holds. In general, the anomalous Green's function $F_{\sigma,\sigma'}^{{\bm p},{\bm k}}(t_1,t_2)$ can contain even and odd components with unequal weight. In this sense, the odd component $F_{o,\sigma,\sigma'}^{{\bm p},{\bm k}}(t_1,t_2)$ can be regarded as the ``hidden order parameter'' of odd-frequency pairing. To the best of our knowledge, no proposal has been made to directly detect $F_{o,\sigma,\sigma'}^{{\bm p},{\bm k}}(t_1,t_2)$ so far.
	
	In this Letter, we suggest a feasible way to directly detect the absolute square of the anomalous Green's function, i.e. $|F_{\sigma,\sigma'}^{{\bm p},{\bm k}}(t_1,t_2)|^2$. If this anomalous Green's function had even and odd frequency components then an exchange of the two time arguments would result in two different signals, proportional to $|F_{e,\sigma,\sigma'}^{{\bm p},{\bm k}}(t_1,t_2) \pm F_{o,\sigma,\sigma'}^{{\bm p},{\bm k}}(t_1,t_2)|^2$. This feature is the mathematical working principle behind our detection scheme, which is physically based on time- and angle-resolved photoelectron fluctuation spectroscopy.
	
	Concretely, we envision a variation of an angle-resolved photoemission spectroscopy (ARPES) setup, as shown in Fig.~\ref{fig:setup}. In a static ARPES protocol, a probe photon beam ejects photoelectrons, which are collected in a momentum-resolving detector. This signal is proportional to the expectation value of the electron number with the wavevector $\vec{p}$ in the scattering state after the light-matter interaction. Here, we instead consider \emph{two detectors} and \emph{two short probe pulses} $A$ and $B$. The two detectors can be viewed as being part of a multi-hit detection scheme, which is able to detect two photoelectrons with momenta $\vec{p}$ and $\vec{p}'$ within a certain time range \cite{kutnyakhov:rsi20}. This gives rise to correlation signals of the form $I^{(2)}_{\vec{p},\sigma;\vec{p}',\sigma'} = \langle n_{\vec{p},\sigma} n_{\vec{p}',\sigma} \rangle$. A similar setup -- with a single probe pulse -- was analyzed by Stahl and Eckstein \cite{stahl:prb19}, who showed that such correlation signals can probe the anomalous Green's function of superconductors. In the single probe pulse case, however, it is not possible to exchange the time arguments $t_1$ and $t_2$ of the anomalous Green's function. As discussed above, this exchange is essential to directly probe odd-frequency pairing. Hence, the crucial difference between the proposal by Stahl and Eckstein and our proposal is the number of probe pulses applied to the sample. We show below that the two probe pulses allow for a time-resolved detection of odd-frequency pairing.
	
	\begin{figure}
		\centerline{\includegraphics[width=0.4\textwidth]{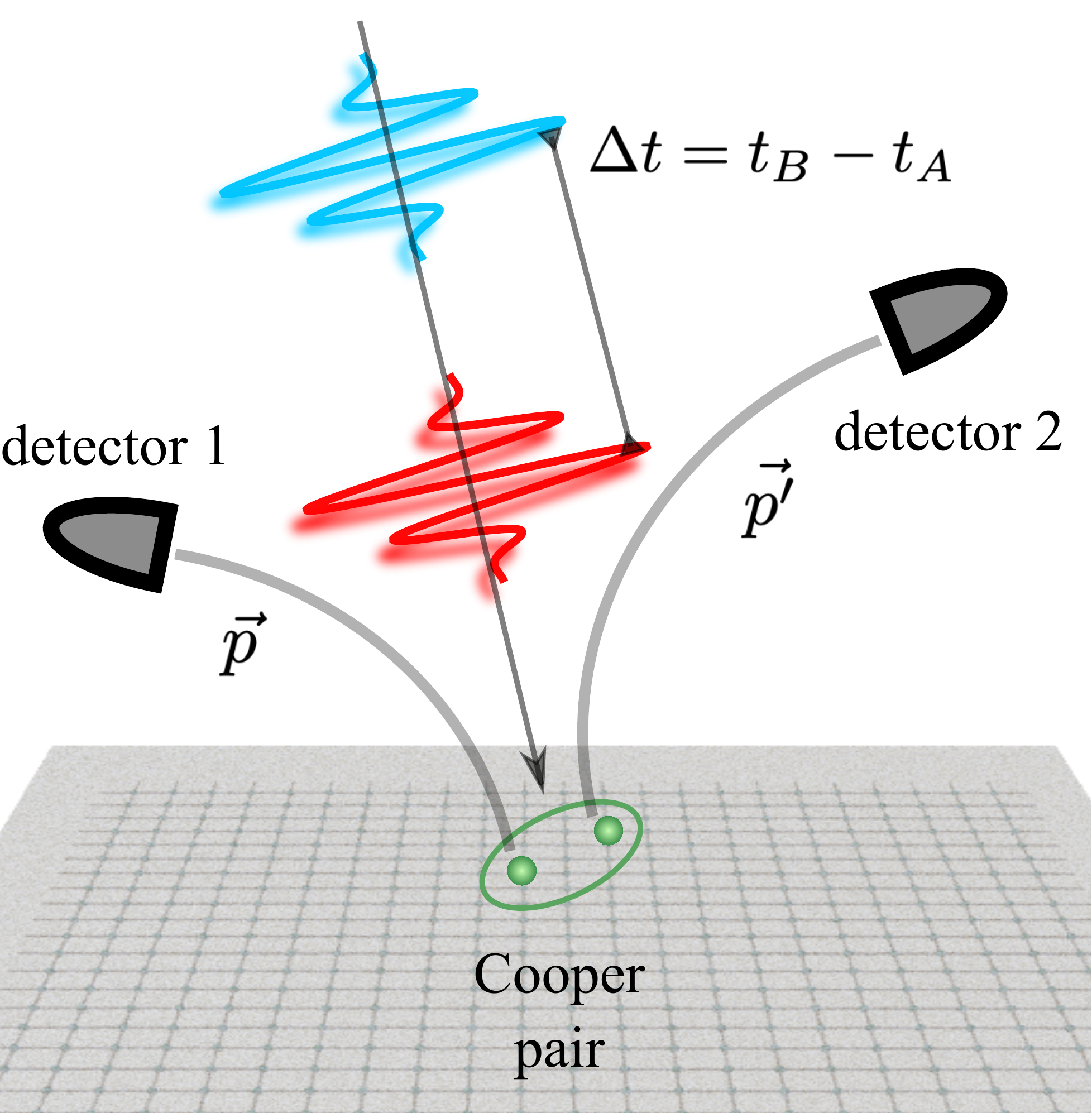}}
		\caption{Schematic picture of the setup. The superconducting sample is subjected to two separate probe pulses $A$ and $B$ with their temporal envelopes centered at times $t_A$ and $t_B$, respectively. One possible detection event is the following one: The electron emitted due to pulse $A$ has momentum ${\bm p}$ and is registered by detector 1, while the electron with momentum ${\bm p'}$ emitted due to the pulse $B$ is registered by detector 2. These two electrons could be correlated if they were initially correlated in the sample.}
		\label{fig:setup}
	\end{figure}
	
	{\it Model.--} The Hamiltonian describing photoemission of electrons due to the two probe pulses $A$ and $B$ can be written as
	\begin{eqnarray}
	H=\sum_{{\bm k},{\bm p},\sigma,\sigma'}S_A^*(t)e^{i\Omega_At}M_{{\bm k},{\bm p}}^{\sigma,\sigma'}a_{{\bm p},\sigma'}^\dagger c_{{\bm k},\sigma}+\nonumber\\+S_B^*(t)e^{i\Omega_Bt}M_{{\bm k},{\bm p}}^{\sigma,\sigma'}b_{{\bm p}\sigma'}^\dagger c_{{\bm k},\sigma}+H.c.,
	\label{Ham:LM}
	\end{eqnarray}
	where $c_{{\bm k},\sigma}$ is an operator of an electron in the material with momentum ${\bm k}$ and spin $\sigma$. Likewise, $a_{{\bm p},\sigma'}$ and $b_{{\bm p},\sigma'}$ are operators of photoemitted electrons with momenta ${\bm p}$ and spins $\sigma'$ via pulse $A$ and pulse $B$, respectively. The matrix element $M_{{\bm k},{\bm p}}^{\sigma\sigma'}$ characterizes emission and is often approximated as $M_{{\bm k},{\bm p}}^{\sigma\sigma'}=M_0\delta_{{\bm k},{\bm p}}\delta_{\sigma\sigma'}$. This approximation is valid if the emitted electrons have the same momentum and spin as they have in the material. The pulses $A$ and $B$ have temporal envelopes $S_A(t)$ and $S_B(t)$ and frequencies $\Omega_A$ and $\Omega_B$, respectively. It is important that the pulses do not overlap. If this was not the case, our protocol for detecting odd-frequency pairing would not work (as further explained below). The Hamiltonian of emitted electrons via pulse $A$ is
	\begin{eqnarray}
	H_{a}=\sum_{{\bm p},\sigma}E_pa^\dagger_{{\bm p},\sigma}a_{{\bm p},\sigma}=\sum_{{\bm p},\sigma}E_pn^a_{{\bm p},\sigma},
	\end{eqnarray}
	and analogously for emitted electrons via pulse $B$. The spectrum of emitted electrons is $E_p=p^2/2m$, where $m$ is the bare mass of an electron.
	
	We consider a material that has superconducting pairing. Hence, the Hamiltonian $H_c$ is assumed to contain terms of the (normal state) kinetic energy of the electron, like $\xi_pc_{{\bm p},\sigma}^\dagger c_{{\bm p},\sigma}$, as well as mean-field terms of the type $\Delta^{p}_{\sigma\sigma'}c^\dagger_{-{\bm p},\sigma'}c^\dagger_{{\bm p},\sigma}$, where $\Delta_{\sigma\sigma'}^{p}$ is the superconducting pairing potential. For now, we do not specify $H_c$ in more detail, because different materials can have different forms of the Hamiltonian that do not affect our general considerations for the detection of odd-frequency pairing significantly. For example, the material may have multiple bands. Then, an additional band index has to be taken into account.
	
	{\it Photoemission signals.--} Following the standard description of electron photoemission, we define the total population of the emitted electrons with momentum ${\bm p}$ and spin ${\bm \sigma}$ as
	\begin{eqnarray}
	I^{(1)}_{{\bm p},\sigma}=\langle n_{{\bm p},\sigma}\rangle_{t=\infty}=\langle \mathcal{S}^\dagger(n^{a}_{{\bm p},\sigma}+n^{b}_{{\bm p},\sigma})\mathcal{S}\rangle_0,
	\end{eqnarray}
	where the averaging $\langle...\rangle_0$ is over the initial state of the system, before applying pulses and in absence of emitted electrons. The time evolution matrix is defined as $\mathcal{S}=Te^{-i\int_{-\infty}^\infty d\tau H(\tau)}$.
	Since we are interested in the total population, the final time is taken to be $t=\infty$. The statistical correlations of photoemission events from all pulses are
	\begin{eqnarray}
	I^{(2)}_{{\bm p},\sigma;{\bm p'},\sigma'}=\langle n_{{\bm p},\sigma}n_{{\bm p'},\sigma'}\rangle_{t=\infty}.
	\end{eqnarray}
	Physically, this describes the correlations of two electrons emitted via two probe photons from the applied pulses. In case ${\bm p}\neq {\bm p'}$, $I^{(2)}_{{\bm p},\sigma;{\bm p'},\sigma'}$ depends on a two-point Green's function with respect to electrons in the material $G_{\sigma_2,\sigma_1,\sigma_2',\sigma_1'}^{{\bm k}_2,{\bm k}_1,{\bm k}_1',{\bm k}_2'}(\tau_2,\tau_1,\tau_2',\tau_1')=\langle \bar{T}[c_{{\bm k}_2,\sigma_2}^\dagger(\tau_2) c_{{\bm k}_1,\sigma_1}^\dagger(\tau_1)]T[c_{{\bm k}_1',\sigma_1'}(\tau_1') c_{{\bm k}_2',\sigma_2'}(\tau_2')]\rangle$, where $\bar{T}$ denotes anti-time ordering, which comes from the expansion of $\mathcal{S}^\dagger$ up to second order (see Section S1 of \cite{SM} for the detailed derivation of the signal). This two-point Green's function can be simplified via Wick's theorem.
	
	Following the description above, the fluctuations of the correlations are defined as
	\begin{eqnarray}
	\Delta I_{{\bm p},\sigma;{\bm p'},\sigma'}=I^{(2)}_{{\bm p},\sigma;{\bm p'},\sigma'}-I_{{\bm p},\sigma}^{(1)}I_{{\bm p'},\sigma'}^{(1)}.
	\end{eqnarray}
	In case of ${\bm p'}=-{\bm p}$, the particle-number-conserving terms of type $\langle c^\dagger c\rangle_0$ cancel and the signal directly depends on the anomalous Green's function via \cite{SM}
	\begin{eqnarray}
	\label{eq:signalDeltaI}
	\Delta I_{{\bm p},\sigma;-{\bm p},\sigma'}=\bigg|\int_{-\infty}^\infty d\tau_1d\tau_2\Xi({\bm p},\tau_1,\tau_2)F_{\sigma',\sigma}^{-{\bm p},{\bm p}}(\tau_1,\tau_2)\bigg|^2,\ \ \ \
	\end{eqnarray}
	where $\Xi({\bm p},\tau_1,\tau_2)$ is a function of pulse shapes, frequencies of applied pulses, $M_0$, and kinetic energy of emitted electrons.
	
	We can deduce from Eq.~(\ref{eq:signalDeltaI}) that, applying a single pulse and detecting emitted electrons in case of $M_{\sigma\sigma'}^{{\bm k},{\bm p}}=M_0\delta_{{\bm k},{\bm p}}\delta_{\sigma,\sigma'}$, this does not enable us to obtain any signal stemming from odd-frequency pairing, because in that case $\Xi({\bm p},\tau_1,\tau_2)=\Xi({\bm p},\tau_2,\tau_1)=M_0^2 S^*(\tau_1)S^*(\tau_2)e^{i(\Omega+E_p)(\tau_1+\tau_2)}$ is an even function with respect to an exchange of the two time arguments. Therefore, the convolution with any odd component of the anomalous Green's function, $F^{-{\bm p},{\bm p}}_{\sigma',\sigma}(\tau_1,\tau_2)$, vanishes.
	
	Thus, in order to detect odd-frequency superconductivity, we need to perform a time-resolved measurement, implying $\Xi({\bm p},\tau_1,\tau_2)\neq\Xi({\bm p},\tau_2,\tau_1)$. To achieve this goal, we suggest to apply two pulses $A$ and $B$ that do not overlap in time. For pedagogical reasons, we first restrict ourselves to a specific process: Detector 1 measures all electrons emitted via pulse $A$ with momentum ${\bm p}$ and spin $\sigma$, while detector 2 measures all electrons emitted via pulse $B$ with momentum $-{\bm p}$ and spin $\sigma'$. (Later, we explain what happens in the more general situation, where all allowed processes are taken into account.) In this specific detection scheme, we only need to analyze the term $\Delta I^{ab}_{{\bm p},\sigma;-{\bm p},\sigma'}=\langle n^a_{{\bm p},\sigma}n^b_{-{\bm p},\sigma'}\rangle_{t=\infty}-\langle n^a_{{\bm p},\sigma}\rangle_{t=\infty}\langle n^b_{-{\bm p},\sigma'}\rangle_{t=\infty}$ with
	\begin{eqnarray}
	\Xi({\bm p},\tau_1,\tau_2)=M_0^2S_B^*(\tau_1)S_A^*(\tau_2)e^{i[\Omega_A\tau_2+\Omega_B\tau_1+E_p(\tau_1+\tau_2)]}.\ \ \ \ \
	\end{eqnarray}
	As the pulses $A$ and $B$ are separated in time, the inequality $\Xi({\bm p},\tau_1,\tau_2)\neq\Xi({\bm p},\tau_2,\tau_1)$ holds if the two photons in pulses $A$ and $B$ are distinguishable, for instance, by energy (or any other means). This inequality allows for a finite signal related to odd-frequency pairing.
	
	The temporal envelopes $S_A(t)$ and $S_B(t)$ are likely to be close to Gaussian shape in real experiments. For simplicity, we assume them to be infinitely short $\delta$ functions, i.e. $S_A(t)=S_0\delta(t-t_A)$ and $S_B(t)=S_0\delta(t-t_B)$ \cite{footnote:delta}. In this case, the signal is
	\begin{eqnarray}
	\label{eq:DeltaIabdelta}
	\Delta I^{ab}_{{\bm p},\sigma;-{\bm p},\sigma'}=(S_0M_0)^4|F_{\sigma',\sigma}^{-{\bm p},{\bm p}}(t_B,t_A)|^2.
	\end{eqnarray}
	Evidently, it can serve as a direct probe to odd-frequency superconductivity: If $F$ has an odd component $F_o$ and an even component $F_e$ then an exchange of the two time arguments in Eq.~(\ref{eq:DeltaIabdelta}) allows us to probe $|F_e+F_o|^2$ and $|F_e-F_o|^2$.
	
	Depending on the characteristics of the materials and detection schemes, we have to take additional terms of the type shown in Eq.~(\ref{eq:DeltaIabdelta}) into account. For instance, we have to sum over spin degrees of freedom if the detection scheme is not spin-resolved. Moreover, we have to sum over terms like $\Delta I^{aa}$, $\Delta I^{bb}$, and $\Delta I^{ba}$ if they happen to yield a finite signal and both detectors are able to detect photoelectrons of any of the two types $A$ and $B$. In the example discussed below, the terms $\Delta I^{aa}$ and $\Delta I^{bb}$ do not depend on the time difference and produce just a constant shift of the overall signal, whereas the term $\Delta I^{ba}$ yields an identical signal to $\Delta I^{ab}$.
	
	{\it Two-band superconductor.--} Below, we demonstrate how our proposal works within a model of a two-band superconductor developed by Parhizgar and Black-Schaffer \cite{parhizgar:arxiv21}. It should be kept in mind that our proposal is generally applicable to any odd-frequency superconductor. The demonstration with respect to the two-band superconductor is just an example, in which many features can be understood analytically. The number of bands matter for the detection scheme. In the two-band case, the energy difference between the bands can be employed to distinguish the two probe pulses. In the single-band case, this is not an option. Then, other degrees of freedom (e.g. spin) have to be taken into account for this purpose.
	
	The Hamiltonian of a two-band superconductor in the basis $\psi_{\bm p}^\dagger=(c_{1,{\bm p},\uparrow}^\dagger,c_{2,{\bm p},\uparrow}^\dagger,c_{1,-{\bm p},\downarrow},c_{2,-{\bm p},\downarrow})^T$ can be written as
	\begin{eqnarray}
	H_c=(\xi_+\tau_0+\xi_-\tau_z+\xi_{12}\tau_x)\gamma_z+(\delta_+\tau_0+\delta_-\tau_z)\gamma_x,
	\end{eqnarray}
	where $\tau_{0,x,z}$ are Pauli matrices in band space $(1,2)$ and $\gamma_{x,z}$ in particle-hole space \cite{footnote:index}. The bands are assumed to be connected by the hybridization $\xi_{12}$, and $\xi_{\pm}=(\xi_1\pm\xi_2)/2$, where $\xi_{1(2)}=t_{1(2)}p^2-\mu-(-1)^{1(2)}\delta\mu/2$. The superconducting pairing is spin-singlet in each band with the order parameters $\delta_{1(2)}=\delta_++(-)\delta_-$. The anomalous Green's function $F$ in frequency space consisting of even- and odd-frequency contributions have been derived in Ref.~\cite{parhizgar:arxiv21}. For our purposes, we need to transfer them to time representation by contour integration with the choice of the poles as $\mp\varepsilon_+\pm i\eta$ and $\mp\varepsilon_-\pm i\eta$ with $\eta\rightarrow+0$ (see the detailed derivation for the time representation of $F$ in Section S2 of \cite{SM}). After Fourier transformation, we obtain for the time difference $t>0$
	\begin{eqnarray}
	\label{eq:Fe}
	\nonumber &&F^{p,-p}_{e,\uparrow\downarrow}(t)=\frac{i}{\varepsilon_-^2-\varepsilon_+^2}\left[\left(\frac{e^{-i\varepsilon_+t}}{\varepsilon_+}-\frac{e^{-i\varepsilon_-t}}{\varepsilon_-}\right)\times\right. \\ &&\left.\nonumber [\xi_{12}(\delta_{+}\xi_{+}-\delta_{-}\xi_{-})\tau_x-\frac{\alpha_{+}}{4}(\tau_0+\tau_z)-\frac{\alpha_{-}}{4}(\tau_0-\tau_z)]+\right.\\
	&&\left.+\frac{1}{2}(\delta_+\tau_0+\delta_-\tau_z)(\varepsilon_+e^{-i\varepsilon_+t}-\varepsilon_-e^{-i\varepsilon_-t})\right],\\
	\label{eq:Fo}
	&&F^{p,-p}_{o,\uparrow\downarrow}(t)=\frac{(e^{-i\varepsilon_+ t}-e^{-i\varepsilon_- t})\delta_-\xi_{12}}{\varepsilon_+^2-\varepsilon_-^2}\tau_y,
	\end{eqnarray}
	where
	$\alpha_\pm=(\xi_+\mp\xi_-)^2(\delta_+\pm\delta_-)+(\delta_+\mp\delta_-)(\xi_{12}^2+\delta_+^2-\delta_-^2)$ and
	$\varepsilon_{\pm}^2=\xi_+^2+\xi_-^2+\xi_{12}^2+\delta_+^2+\delta_-^2\pm2\sqrt{(\delta_+\delta_-+\xi_+\xi_-)^2+\xi_{12}^2(\delta_-^2+\xi_+^2)}$. Note that these Green's functions depend on the time difference only.
	
	From Eq.~(\ref{eq:Fo}), it follows that the applied pulses must address different electron bands. Otherwise, the signal from the odd-frequency pairing does not contribute to the photoelectron fluctuation signal. This requirement should be feasible, if the gap between the bands is much wider than the pulse width. Then, we can adjust $\Omega_A=E_p+E_{1,{\rm exit}}$, where $E_{1,{\rm exit}}$ constitutes the energy necessary for an electron stemming from band 1 to reach the vacuum energy level, and, likewise, $\Omega_B=E_p+E_{2,{\rm exit}}$ for an electron stemming from band 2.
	
	{\it Photoelectron fluctuations for two-band model.--} In Fig. \ref{fig:signal}, we plot the dependence of the full signal, $\Delta I^{ab}_{{\bm p},-{\bm p}}/\tilde{I}=(1/2)(|F^{-{\bm p},{\bm p}}_{\uparrow\downarrow}(t_B-t_A,2,1)|^2+|F^{-{\bm p},{\bm p}}_{\downarrow\uparrow}(t_B-t_A,2,1)|^2)$ with $\tilde{I}=2(S_0M_0)^4$ as well as its odd- ($|F_o|^2$) and even-frequency ($|F_e|^2$) components with respect to the dimensionless momentum $p/\tilde{p}$ for shorter (a) and longer (b) time differences between the pulses, $t_B-t_A$. For this model, the following symmetry holds $F_{\sigma,\sigma'}=-F_{-\sigma,-\sigma'}$. Therefore, the two terms in $\Delta I^{ab}_{{\bm p},-{\bm p}}$ are identical. We define the units of energy in such a way that $t_{1}\tilde{p}^2=1$. Moreover, time differences are shown in units of inverse energy. For concreteness, we use the following parameters: $t_2\tilde{p}^2=0.5$, $\delta_+=0.03$, $\delta_-=0.02$, $\xi_{12}=0.05$, $\mu=0.3$, $\delta\mu=-0.2$, which correspond to two electron-like bands.
	\begin{figure}
		\centerline{\includegraphics[width=0.45\textwidth]{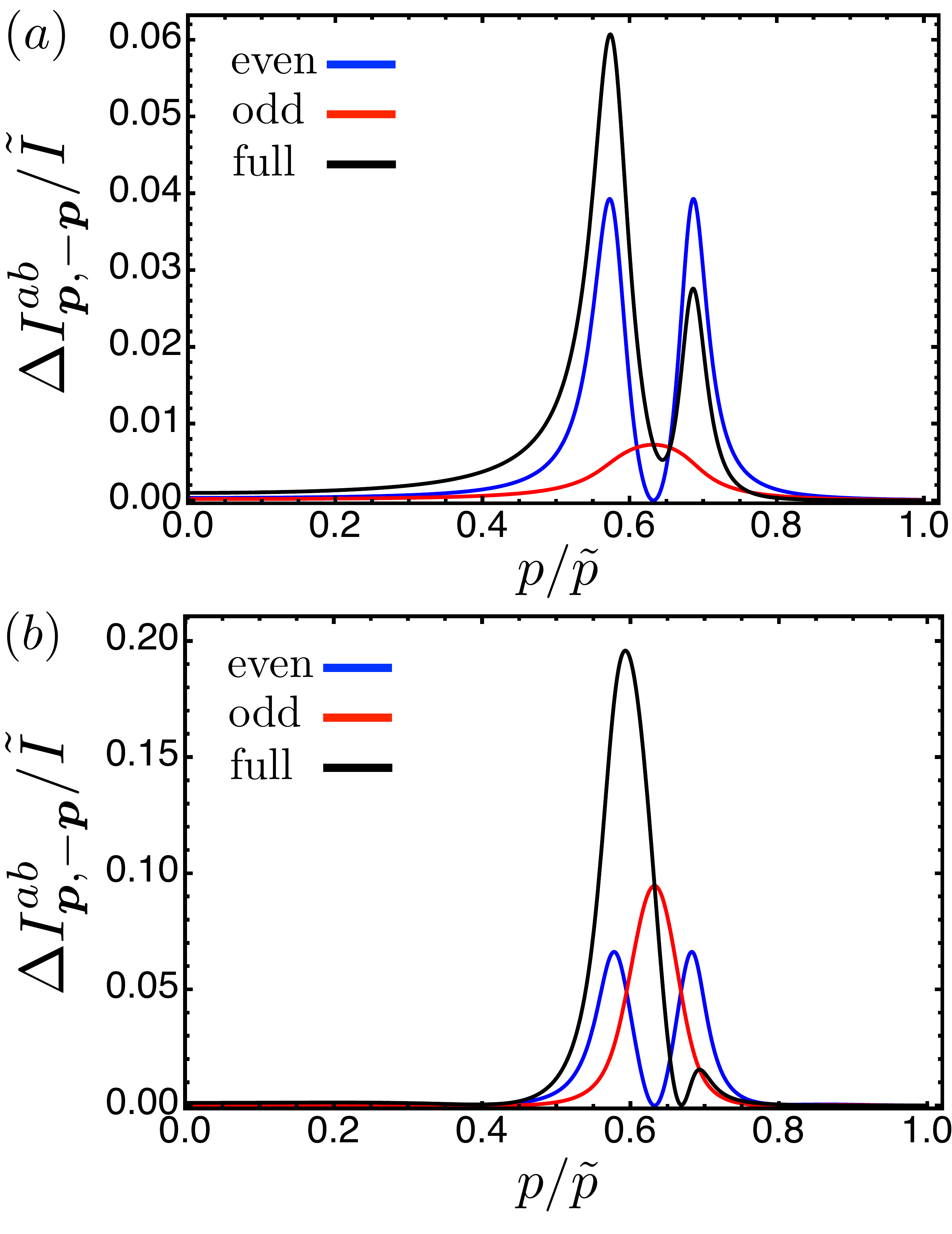}}
		\caption{Momentum dependence of odd (red), even (blue), and full signal (black) for the model of a two-band superconductor for (a) $t_B-t_A=10$ and (b) $t_B-t_A=40$. The odd part of the signal has one peak, while the even part has two peaks. }
		\label{fig:signal}
	\end{figure}
	In Fig.~\ref{fig:signal}(a), we consider a shorter time difference, $t_B-t_A=10$, than in Fig.~\ref{fig:signal}(b), for which we take $t_B-t_A=40$. In Fig.~\ref{fig:signal}(a), the signal consists of two peaks, because the contribution from the even-frequency part is dominating over the odd-frequency one. However, in Fig.~\ref{fig:signal}(b), the signal becomes close to a single peak, as the odd-frequency contribution starts to dominate. For $t_B-t_A\rightarrow 0$, the contribution from the odd-frequency part is vanishing as it should.
	This analysis demonstrates that the content of the odd- and the even-frequency contribution to superconducting pairing can vary as a function of the time argument of the anomalous Green's function as well as the momenta that appear in it. Note that given momenta fix the corresponding energy. The plots shown in Fig.~\ref{fig:signal}(a,b) are model-dependent but we expect that similar features appear in other models as well as real materials.
	%
	%
	%
	\begin{figure}
		\centerline{\includegraphics[width=0.45\textwidth]{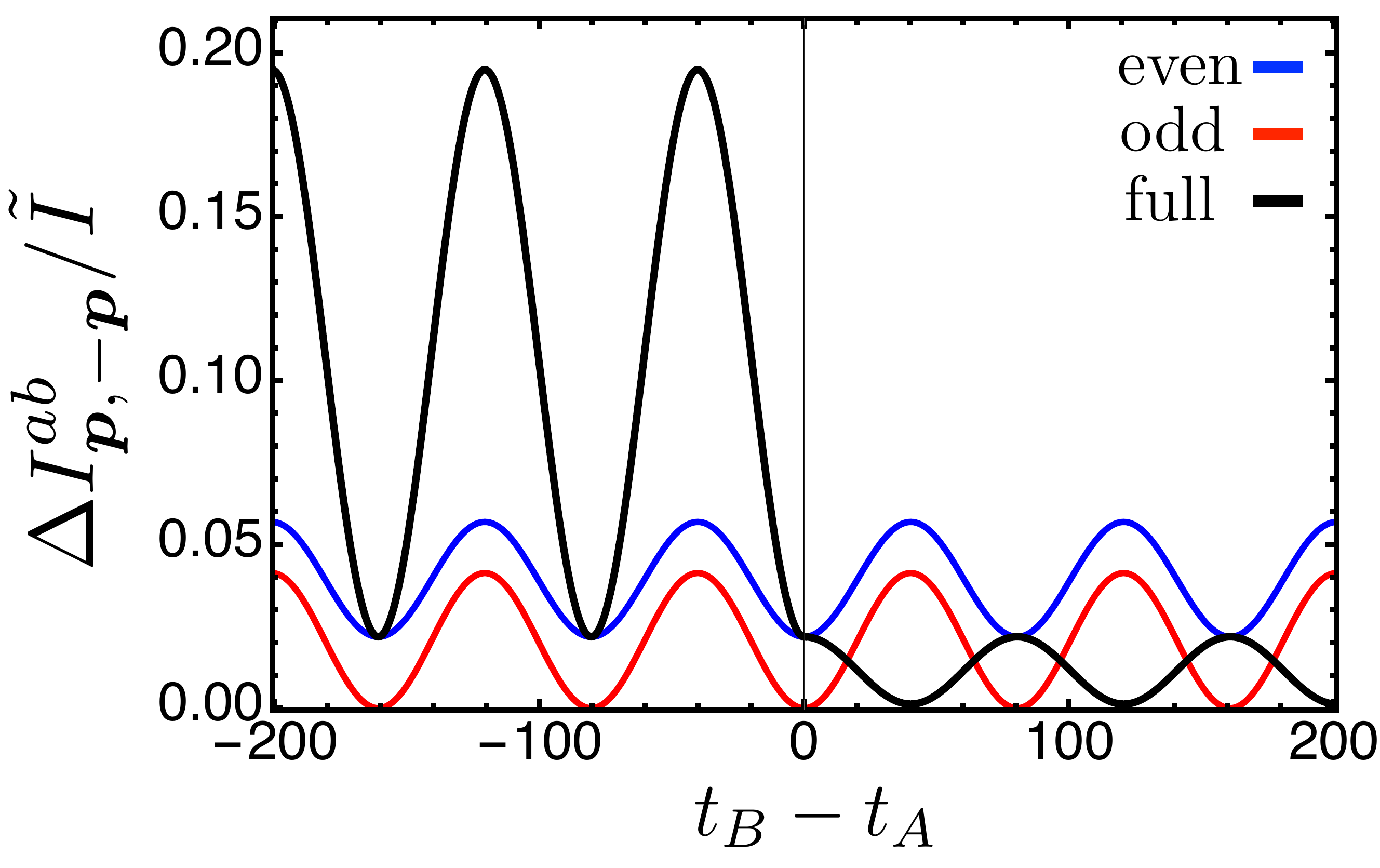}}
		\caption{Time dependence of odd (red), even (blue), and full signal (black) for the model of a two-band superconductor at $p=p_{\rm max}+\delta p$, $\delta p/\tilde{p}=0.04$. The full signal is different for a different ordering of pulses $A$ and $B$ since they address different bands. The odd component changes its sign with the order of the band indices, while the even one does not. }
		\label{fig:SignalTime}
	\end{figure}
	
	In Fig. \ref{fig:SignalTime}, we plot the dependence of the full signal, $\Delta I^{ab}_{{\bm p},-{\bm p}}$, as well as its even and odd components with respect to the time difference between the pulses $t_B-t_A$ at a given momentum. Negative time difference $t_B-t_A$ means that the order of the pulses is exchanged.
	Thus, we obtain the signal $\Delta I^{ab}_{{\bm p},-{\bm p}}/\tilde{I}=|F^{-{\bm p},{\bm p}}_{o,\uparrow,\downarrow}(t_B-t_A,2,1)+F^{-{\bm p},{\bm p}}_{e,\uparrow,\downarrow}(t_B-t_A,2,1)|^2$ for $t_B>t_A$ and $\Delta I^{ab}_{{\bm p},-{\bm p}}/\tilde{I}=|F^{{\bm p},-{\bm p}}_{o,\downarrow,\uparrow}(t_A-t_B,1,2)+F^{{\bm p},-{\bm p}}_{e,\downarrow,\uparrow}(t_A-t_B,1,2)|^2$ for $t_A>t_B$. We stress that for the given two-band model the exchange of spin and momentum indices does not change $\Delta I^{ab}_{{\bm p},-{\bm p}}$. Only the exchange of band indices provides an additional minus sign due to $F_o(1,2)=-F_o(2,1)$, yielding different signal for $t_B-t_A>0$ and $t_B-t_A<0$.
	
	 Note that Fig. \ref{fig:SignalTime} illustrates our main result. It shows that the coexistence of even- and odd-frequency pairing results in an asymmetry of the photoelectron fluctuation signal with respect to the exchange of the two pulses. This result is by no means specific to the two-band model under consideration. All it needs for the asymmetry in the photoelectron fluctuation signal to appear is a significant contribution of even- and odd-frequency pairing amplitudes at a certain momentum. A variety of materials fulfills this criterion. Examples thereof are: MgB$_2$ \cite{aperis:prb2015}, Rashba wires on superconducting substrates \cite{ebisu:prb2015,takagi:prb2020}, Sr$_2$RuO$_4$ \cite{komendova:prl2017}, iron-pnictide superconductors \cite{nica:npj2017}, doped Bi$_2$Se$_3$ \cite{parhizgar:scirep2017,schmidt:prb2020}, UPt$_3$ \cite{triola:prb2018}, dilute magnetic superconductors \cite{santos:prr2020}, and superconducting Weyl semimetals \cite{parhizgar:npj2020,dutta:prb2020}.

	{\it Experimental issues.--} Our proposal for detecting time-nonlocal pairing via photoelectron fluctuation spectroscopy is based on the following key ingredients: (i) two short probe pulses that both photoemit electrons and ideally do not overlap in time; (ii) adjustable probe photon energies that differ by the energetic separation of the Bogolyubov bands in a given multiband superconductor  \cite{footnote:multiband}; (iii) a detector that measures multi-hit signals of photoelectrons at opposite momenta. Point (i) is related to the time resolution of state-of-the-art ARPES setups, set by the minimal duration of laser pulses on the order of few tens of femtoseconds \cite{torre:arxiv21}. This time resolution puts a lower bound on features in the time-dependent signal presented in Fig.~\ref{fig:SignalTime} that can be resolved  in an experiment. Importantly, while the temporal oscillations might be washed out if their period is smaller than the time resolution of the experiment, the salient features of odd-frequency pairing should still be observable even in that case. Point (ii) puts a practical limitation on the energetic separation required in the multi-band superconductor. In a probe pulse of short duration, the Heisenberg uncertainty principle imposes a finite energy bandwidth of a photon pulse via $\delta t \delta E \geq \hbar \approx 658$ meV fs. For a probe duration as short as 30 fs, this imposes a requirement that the Bogolyubov bands are separated by 20 meV or more at the relevant probe momentum points near the Fermi momentum. We note that this criterion is fulfilled in a variety of multiband superconductors, such as iron pnictides \cite{nica:npj2017} or Sr$_2$RuO$_4$ \cite{baumberger}. Taken together, conditions (i) and (ii) suggest that probe photon energies and probe durations need to be carefully adjusted to the superconductor under consideration. Point (iii) poses a practical challenge to photoelectron detector technologies that are presently available \cite{sobota:rmp21}. A similar measurement where a single pulse ejects two photoelectrons (double photoemission) has been demonstrated \cite{schumann:prl07,schumann:prb06,trutzschler:prl17,chiang:pss20}. For our proposal of angle-resolved photoelectron fluctuation spectroscopy, the development of momentum microscopes with multi-channel detectors that allow to measure multi-hit signals appears promising \cite{kutnyakhov:rsi20}.

	{\it Summary.--} In conclusion, we have suggested a measurement scheme based on time- and angle-resolved photoelectron
	fluctuation spectroscopy for the direct detection of odd-frequency superconductivity, a long-standing problem in the field of unconventional superconductivity. We have applied our scheme to a two-band superconductor, analyzed the corresponding signal, and suggested a way to extract the odd-frequency part of it.
	
	\begin{acknowledgments}
		We acknowledge useful discussions with Alexander Balatsky, Hendrik Bentmann, Tero Heikkil\"a, Kirill Nagaev, and Laurenz Rettig. This work was supported by the DFG (SPP1666 and SFB1170 ``ToCoTronics''), the W{\"u}rzburg-Dresden Cluster
		of Excellence ct.qmat, EXC2147, project-id 390858490, and the Elitenetzwerk Bayern Graduate School on ``Topological Insulators''. M.A.S.~acknowledges financial support through the DFG Emmy Noether program (SE 2558/2).
		F.S. acknowledges support from the Cluster of Excellence 'Advanced Imaging of Matter' of the Deutsche Forschungsgemeinschaft (DFG) - EXC 2056 - project ID 390715994.
	\end{acknowledgments}
\begin{widetext}
\section*{Supplemental Material }

\maketitle
\renewcommand{\theequation}{S\arabic{equation}}
\setcounter{equation}{0}
\renewcommand{\thefigure}{S\arabic{figure}}
\renewcommand{\figurename}{Supplementary Fig.}

\setcounter{figure}{0}
\renewcommand{\thesection}{S\arabic{section}}
\setcounter{section}{0}

In this Supplemental Material, we present additional details and calculations regarding: 1) the derivation of the signal in an experiment with two probe pulses and 2) the derivation of the time representation of anomalous Green functions for the two-band superconductor model. 

\section{S1. Derivation of the signal in an experiment with two probe pulses}
The derivation for the case of a single pulse was discussed in Ref. \onlinecite{stahl:prb19}. Here, we describe our derivation for the two non-overlapping pulses. The Hamiltonian describing interactions between the pulses and the studied material is
\begin{eqnarray}
\label{eq:Htwopulses}
H=\sum_{{\bm k},{\bm p},\sigma,\sigma'}S_A(t)^*e^{i\Omega_A t}M^{\sigma,\sigma'}_{{\bm k},{\bm p}}a^\dagger_{{\bm p}\sigma'}c_{{\bm k},\sigma}+S_B(t)^*e^{i\Omega_B t}M^{\sigma,\sigma'}_{{\bm k},{\bm p}}b^\dagger_{{\bm p}\sigma'}c_{{\bm k},\sigma}+H. c.
\end{eqnarray}
The pulse envelope functions $S_A(t)$ and $S_B(t)$ have finite widths and no overlap. Therefore, we assume that $\langle a_{{\bm p},\sigma}b^\dagger_{{\bm p},\sigma}\rangle_0=\langle b_{{\bm p},\sigma}a^\dagger_{{\bm p},\sigma}\rangle_0=0$. Hence, the vacuum state of emitted electrons can be described as a product of vacuum states for electrons emitted via pulse $A$ and electrons emitted via pulse $B$: $|0\rangle_{\rm emit}=|0\rangle_a|0\rangle_b$. 

The total population of the outgoing state is
\begin{eqnarray}
I_{{\bm p},\sigma}^{(1)}=\langle n^a_{{\bm p},\sigma}+n^b_{{\bm p},\sigma}\rangle_{t=\infty}.
\end{eqnarray}
The statistical correlation of photoemission events is
\begin{eqnarray}
I^{(2)}_{{\bm p},\sigma;{\bm p'},\sigma'}=\langle (n^a_{{\bm p},\sigma}+n^b_{{\bm p},\sigma})(n^a_{{\bm p'},\sigma'}+n^b_{{\bm p'}\sigma'})\rangle_{t=\infty}. 
\end{eqnarray}
In terms of the time evolution matrix, $\mathcal{S}=T\exp{\left[-i\int_{-\infty}^\infty H(\tau)d\tau\right]}$ with $T$ denoting time ordering, we obtain
\begin{eqnarray}
I^{(1)}_{{\bm p},\sigma}&=&\langle \mathcal{S}^\dagger (n^a_{{\bm p},\sigma}+n^b_{{\bm p},\sigma})\mathcal{S}\rangle_0=\langle \mathcal{S}_a^\dagger n^a_{{\bm p},\sigma} \mathcal{S}_a\rangle_0+\langle \mathcal{S}_b^\dagger n^b_{{\bm p},\sigma}\mathcal{S}_b\rangle_0,\\
I^{(2)}_{{\bm p},\sigma;{\bm p'},\sigma'}&=&\langle \mathcal{S}^\dagger(n^a_{{\bm p},\sigma}+n^b_{{\bm p},\sigma})(n^a_{{\bm p'},\sigma'}+n^b_{{\bm p'},\sigma'}) \mathcal{S}\rangle_0=\langle \mathcal{S}^\dagger(n^a_{{\bm p},\sigma}n^a_{{\bm p'},\sigma'}+n^a_{{\bm p},\sigma}n^b_{{\bm p'},\sigma'}+n^b_{{\bm p},\sigma}n^a_{{\bm p'},\sigma'}+n^b_{{\bm p},\sigma}n^b_{{\bm p'},\sigma'}) \mathcal{S}\rangle_0.\end{eqnarray}
The first and the last terms in $I^{(2)}_{{\bm p},\sigma;{\bm p'},\sigma'}$ are the same as in Ref. \onlinecite{stahl:prb19}. Therefore we will consider only the second and the third terms here. For the case of pulse $B$ being later than pulse $A$, the second term is
 \begin{eqnarray}
 I^{(2),ab}_{{\bm p},\sigma;{\bm p'},\sigma'}=\frac{1}{4}\langle\int_{-\infty}^\infty d\tau_1 d\tau_2 \bar{T}[H(\tau_1)H(\tau_2)]n_{p\sigma}^an_{p'\sigma'}^b\int_{-\infty}^\infty d\tau_1' d\tau_2' T[H(\tau_1')H(\tau_2')]\rangle_0=\ \ \ \\ \nonumber=\int_{-\infty}^\infty d\tau_1 d\tau_2d\tau_1' d\tau_2'  \sum_{\substack{k_1,p_1,\sigma_1,\gamma_1,\\ k_1',p_1',\sigma_1',\gamma_1',\\k_2,p_2,\sigma_2,\gamma_2,\\k_2',p_2',\sigma_2',\gamma_2'}}M_{k_2,p_2}^{\sigma_2,\gamma_2}M_{k_1,p_1}^{\sigma_1,\gamma_1}M_{k_1',p_1'}^{\sigma_1',\gamma_1'}M_{k_2',p_2'}^{\sigma_2',\gamma_2'}\langle c^\dagger_{k_2,\sigma_2}c^\dagger_{k_1,\sigma_1}c_{k_1',\sigma_1'}c_{k_2',\sigma_2'}\rangle_0\\ \nonumber
S_B(\tau_1)e^{-i\Omega_B\tau_1}S_A(\tau_2)e^{-i\Omega_A\tau_2}S_B(\tau_1')^*e^{i\Omega_B\tau_1'}S_A^*(\tau_2')e^{i\Omega_A\tau_2'}\langle b_{p_1\gamma_1}b_{p'\sigma'}^\dagger\rangle\langle b_{p'\sigma'}b_{p_1'\gamma_1'}^\dagger\rangle \langle a_{p_2\gamma_2}a_{p\sigma}^\dagger\rangle\langle a_{p\sigma} a_{p_2'\gamma_2'}^\dagger\rangle=\\ \nonumber=\int_{-\infty}^\infty d\tau_1 d\tau_2d\tau_1' d\tau_2'  M_0^4\\ \nonumber S_B(\tau_1)S_A(\tau_2)S_B^*(\tau_1')S_A^*(\tau_2')e^{-i\Omega_B\tau_1-i\Omega_A\tau_2+i\Omega_B\tau_1'+i\Omega_A\tau_2'+iE_{p'}(\tau_1'-\tau_1)+iE_p(\tau_2'-\tau_2)}\langle c^\dagger_{p,\sigma}(\tau_2)c^\dagger_{p',\sigma'}(\tau_1)c_{p',\sigma'}(\tau_1')c_{p,\sigma}(\tau_2')\rangle_0.
\end{eqnarray} 
The symbol $\bar{T}$ denotes anti-time ordering due to the fact that we used $\mathcal{S}^\dagger$. The last equality sign was taken assuming $M_{{\bm k},{\bm p}}^{\sigma,\sigma'}=M_0\delta_{{\bm k},{\bm p}}\delta_{\sigma,\sigma'}$. Now, we can reintroduce time-ordering and obtain a two-point Green function
\begin{eqnarray}
G^{{\bm p}, {\bm p'},{\bm p'},{\bm p}}_{\sigma,\sigma',\sigma',\sigma}(\tau_2,\tau_1,\tau_1',\tau_2')=\langle \bar{T}[c^\dagger_{{\bm p},\sigma}(\tau_2)c^\dagger_{{\bm p'},\sigma'}(\tau_1)]T[c_{{\bm p'},\sigma'}(\tau_1')c_{{\bm p},\sigma}(\tau_2')]\rangle_0.
\end{eqnarray}
We can expand it using Wick's theorem:
\begin{eqnarray}
G^{{\bm p}, {\bm p'},{\bm p'},{\bm p}}_{\sigma,\sigma',\sigma',\sigma}(\tau_2,\tau_1,\tau_1',\tau_2')=\langle \bar{T}[c^\dagger_{{\bm p},\sigma}(\tau_2)c^\dagger_{{\bm p'},\sigma'}(\tau_1)]T[c_{{\bm p'},\sigma'}(\tau_1')c_{{\bm p},\sigma}(\tau_2')]\rangle_0=\\ \nonumber=\langle\bar{T}[c^\dagger_{{\bm p},\sigma}(\tau_2)c^\dagger_{{\bm p'},\sigma'}(\tau_1)]\rangle_0\langle T[c_{{\bm p'},\sigma'}(\tau_1')c_{{\bm p},\sigma}(\tau_2')]\rangle_0-\langle c_{{\bm p},\sigma}^\dagger(\tau_2)c_{{\bm p'},\sigma'}(\tau_1')\rangle_0\langle c^\dagger_{{\bm p'},\sigma'}(\tau_1)c_{{\bm p},\sigma}(\tau_2')\rangle_0+\\ \nonumber+\langle c^\dagger_{{\bm p},\sigma}(\tau_2)c_{{\bm p},\sigma}(\tau_2')\rangle_0\langle c^\dagger_{{\bm p'},\sigma'}(\tau_1)c_{{\bm p'},\sigma'}(\tau_1')\rangle_0.
\end{eqnarray}
If the Hamiltonian describing our material does not contain terms proportional to $c^\dagger_{{\bm p},\sigma}c_{{\bm p'},\sigma'}$ for ${\bm p}\neq {\bm p'}$ and $\sigma\neq \sigma'$, which is often the case for a clean material, the second term is zero in case of ${\bm p'}=-{\bm p}$. The last term is a product of two one-point Green functions. 

Thus, for the fluctuations $\Delta I^{ab}_{{\bm p},\sigma;-{\bm p},\sigma'}=I^{(2),ab}_{{\bm p},\sigma;-{\bm p},\sigma'}-I^{(1),a}_{{\bm p},\sigma}I^{(1),b}_{-{\bm p},\sigma'}=\langle n^a_{{\bm p},\sigma}n^b_{-{\bm p},\sigma'}\rangle_{t=\infty}-\langle n^a_{{\bm p},\sigma}\rangle_{t=\infty}\langle n^b_{-{\bm p},\sigma'}\rangle_{t=\infty}$, we obtain
\begin{eqnarray}
\Delta I^{ab}_{{\bm p},\sigma;-{\bm p},\sigma'}=\bigg|\int_{-\infty}^\infty d\tau_1 d\tau_2  M_0^2S_B^*(\tau_1)S_A^*(\tau_2)e^{i\Omega_B\tau_1+i\Omega_A\tau_2+iE_{p}(\tau_1+\tau_2)}\langle Tc_{-{\bm p},\sigma'}(\tau_1)c_{{\bm p},\sigma}(\tau_2)\rangle_0\bigg|^2.
\end{eqnarray}  
In case of a single pulse, the signal has the same form (with all indices $A$ and $B$ being absent)\cite{stahl:prb19}, even though the derivation is slightly different due to the fact that all emitted electrons have the same vacuum state. In Eqs. (6) and (7) and the related discussion in the main text, we present the above equation. 

In case of delta-function-shaped envelopes of pulses, $S_A(t)=S_0\delta(t-t_A)$, $S_B(t)=S_0\delta(t-t_B)$, the signal is simplified into
\begin{eqnarray}
\Delta I^{ab}_{{\bm p},\sigma;-{\bm p},\sigma'}=M_0^4S_0^4\langle c_{-{\bm p},\sigma'}(t_B)c_{{\bm p},\sigma}(t_A)\rangle_0=M_0^4S_0^4 |F_{\sigma',\sigma}^{-{\bm p},{\bm p}}(t_B,t_A)|^2.
\end{eqnarray}
This is Eq. (8) from the main text. If operators $c$ have an additional index, e.g. a band index, the anomalous Green function also acquires indices. In the main text, we consider an example of a two-band superconductor, when pulse $A$ affects band 1 and pulse $B$ affects band 2. Therefore, the signal is
\begin{eqnarray}
\Delta I^{ab}_{{\bm p},\sigma;-{\bm p},\sigma'}=M_0^4S_0^4\langle c_{2,-{\bm p},\sigma'}(t_B)c_{1,{\bm p},\sigma}(t_A)\rangle_0=M_0^4S_0^4 |F_{\sigma',\sigma}^{-{\bm p},{\bm p}}(t_B,t_A, 2,1)|^2.
\end{eqnarray} 
The fact that the pulses address different bands provides interesting results for the dependence of the signal on the time difference between pulses, see Fig. 3 of the main text.

\section{S2. Time dependence of anomalous Green functions of a two-band superconductor}
In this section, we present the way we perform the Fourier transformation of the anomalous Green functions, even and odd, for a model of a two-band superconductor described in Ref. \onlinecite{parhizgar:arxiv21}.

To obtain anomalous Green functions in time representation, we need to calculate $\int_{-\infty}^\infty F(\omega) e^{i\omega t}d\omega$, however, this integral has divergencies at $\pm\varepsilon_{\pm}$, see Eqs. (5) and (6) in Ref. \onlinecite{parhizgar:arxiv21}. Therefore, we employ the circling of the poles analogously to the Landau theorem \cite{landau:jetp58, gorkov:jetp58}. As the Landau theorem applies to a single type of fermions, while we have two bands and thus a combination of their contributions into the Green function, we use it for the denominators respective to different eigenbands:
\begin{eqnarray}
D=(\omega^2-\varepsilon_-^2)(\omega^2-\varepsilon_+^2)=(\omega-\varepsilon_-+i\eta)(\omega+\varepsilon_--i\eta)(\omega-\varepsilon_++i\eta)(\omega+\varepsilon_+-i\eta).
\end{eqnarray}
Thus, the poles are at $\omega_{1,2}=\mp\varepsilon_+\pm i\eta$ and $\omega_{3,4}=\mp\varepsilon_-\pm i\eta$. 
Then, we define the contour in the complex plane $\Omega=\{\omega,i\omega'\}$, see Fig. \ref{fig:contour}. 
\begin{figure}
	\centerline{\includegraphics[width=0.5\textwidth]{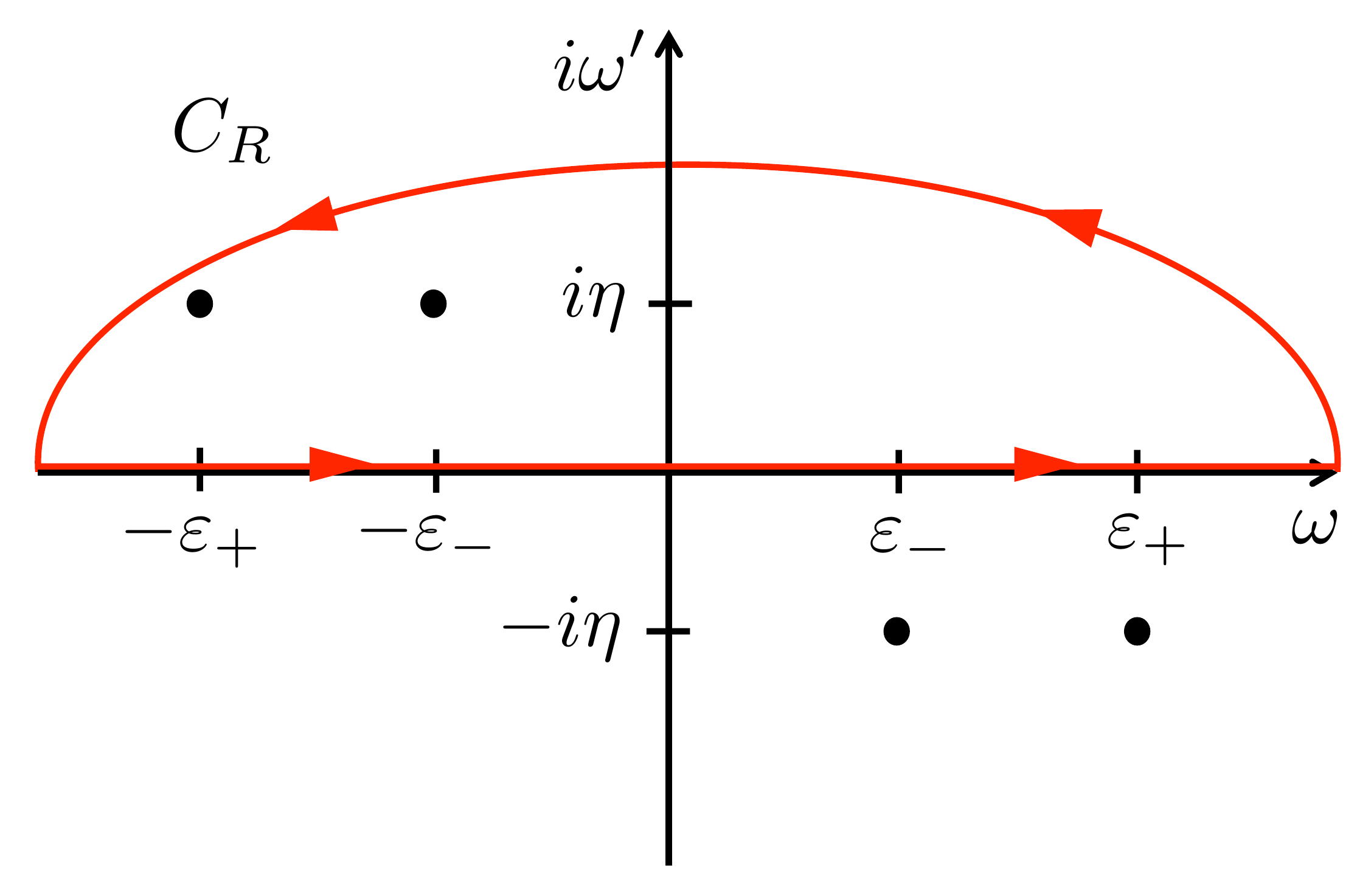}}
	\caption{The contour for integration over frequency.}
	\label{fig:contour}
\end{figure}

The contour has radius $R$ and is defined for the upper half plane. We obtain
\begin{eqnarray}
\int_CF(\Omega)e^{i\Omega t}d\Omega=\int_{-R}^{R} F(\Omega)e^{i\Omega t}d\Omega+\int_{C_{R}} F(\Omega)e^{i\Omega t}d\Omega=2\pi i \sum_{\omega_1,\omega_3} {\rm res}[Fe^{i\Omega t}].
\end{eqnarray}
Let us evaluate the second integral (see e.g. Ref. \onlinecite{polovinkin:book99}). First of all, for large enough $|\Omega|>R_0$,
\begin{eqnarray}
F(\Omega)=\Omega^{-2}(1+O(1/\Omega)).
\end{eqnarray}
We can also find such $R_1\geq R_0$ that at all $|\Omega|\geq R_1$ we have $|1+O(1/\Omega)|\leq 2$, i.e. $|F(\Omega)|\leq 2|\Omega|^{-2}$. Moreover, 
\begin{eqnarray}
|e^{i\Omega t}|=|e^{i(\omega+i\omega')t}|=e^{-\omega't}=e^{-t R\sin{\phi}}.
\end{eqnarray}
This equality holds for $t>0$. Note that $\sin{\phi}\geq \frac{2}{\pi}\phi$ for $\phi\in [0,\pi/2]$. Thus for $R\geq R_1$ we have
\begin{eqnarray}
\left|\int_{C_{R}} F(\Omega)e^{i\Omega t}d\Omega\right|\leq\int_{C_R}2|\Omega|^{-2}e^{-t R\sin{\phi}}|d\Omega|\leq \\ \nonumber 2R^{-2}R\int_0^\pi e^{-\tau R\sin{\phi}} d\phi\leq2R^{-2} R2\int_0^{\pi/2}e^{-2\tau R\phi/\pi}d\phi\leq 2R^{-2} \frac{\pi}{\tau}\rightarrow_{R\rightarrow\infty}0.
\end{eqnarray}
In such a way we obtain
\begin{eqnarray}
\int_{-\infty}^{\infty} F(\Omega)e^{i\Omega t}d\Omega=2\pi i \sum_{\omega_1,\omega_3} {\rm res}[Fe^{i\Omega t}].
\end{eqnarray}
Next we calculate the residues. As these are simply poles of the first order, we obtain for the even component of the signal, $F_e$,
\begin{eqnarray}
{\rm res}_{\omega_1}[F_e(\Omega)e^{i\Omega t}]=\frac{e^{i\omega_1 t}}{(\omega_1-\varepsilon_+)(\omega_1^2-\varepsilon_-^2)}\begin{pmatrix}(\delta_++\delta_-)\omega_1^2-\alpha_+& 2\xi_{12}(\delta_+\xi_+-\delta_-\xi_-) \\ 2\xi_{12}(\delta_+\xi_+-\delta_-\xi_-) & (\delta_+-\delta_-)\omega_1^2-\alpha_-\end{pmatrix},\\
{\rm res}_{\omega_3}[F_e(\Omega)e^{i\Omega t}]=\frac{e^{i\omega_3 t}}{(\omega_3-\varepsilon_-)(\omega_3^2-\varepsilon_+^2)}\begin{pmatrix}(\delta_++\delta_-)\omega_3^2-\alpha_+& 2\xi_{12}(\delta_+\xi_+-\delta_-\xi_-) \\ 2\xi_{12}(\delta_+\xi_+-\delta_-\xi_-) & (\delta_+-\delta_-)\omega_3^2-\alpha_-\end{pmatrix}.
\end{eqnarray}
Then we take a limit $\eta\rightarrow 0$, and obtain:
\begin{eqnarray}
\label{eq:supplFetime}
F_e(t)=\frac{1}{2\pi}\int_{-\infty}^{\infty} F_e(\Omega)e^{i\Omega t}d\Omega=\frac{i}{2(\varepsilon_-^2-\varepsilon_+^2)} \left[\frac{e^{-i\varepsilon_+ t}}{\varepsilon_+}\begin{pmatrix}(\delta_++\delta_-)\varepsilon_+^2-\alpha_+& 2\xi_{12}(\delta_+\xi_+-\delta_-\xi_-) \\ 2\xi_{12}(\delta_+\xi_+-\delta_-\xi_-) & (\delta_+-\delta_-)\varepsilon_+^2-\alpha_-\end{pmatrix}-\right.\\ \nonumber\left.-\frac{e^{-i\varepsilon_- t}}{\varepsilon_-}\begin{pmatrix}(\delta_++\delta_-)\varepsilon_-^2-\alpha_+& 2\xi_{12}(\delta_+\xi_+-\delta_-\xi_-) \\ 2\xi_{12}(\delta_+\xi_+-\delta_-\xi_-) & (\delta_+-\delta_-)\varepsilon_-^2-\alpha_-\end{pmatrix}\right].
\end{eqnarray}
Analogously, for the odd component, $F_o(t)$, we obtain
\begin{eqnarray}
\label{eq:supplFotime}
F_o(t)=\frac{1}{2\pi}\int_{-\infty}^\infty F_o(\Omega)e^{i\Omega t}d\Omega=\frac{i(e^{-i\varepsilon_+ t}-e^{-i\varepsilon_- t})}{\varepsilon_+^2-\varepsilon_-^2}\begin{pmatrix}0& -\delta_-\xi_{12} \\ \delta_-\xi_{12} & 0\end{pmatrix}.
\end{eqnarray}
Note that Eqs. (\ref{eq:supplFetime}) and (\ref{eq:supplFotime}) correspond to Eqs. (10) and (11) of the main text. We simplify the indices of the anomalous Green functions here for ease of notation, keeping only time and even/odd notation.

\end{widetext}


\begin{thebibliography}{}
\bibitem{footnote:delta} This assumption reflects an experiment, in which the temporal line width of the pulse is the shortest time scale of the problem. It should be kept in mind that a small line width might cause problems in the detection scheme if too many energy levels are excited due to energy-time uncertainty. In this case, the line width should be adapted to the problem. However, the principle of our detection scheme does not rely on this assumption.
\bibitem{footnote:splitting}{The time at which the splitting happens can be estimated as half of the time necessary to reach zero of $|F_o|^2$ at $p_{\rm max}$, i.e. $t_{\rm sp}=T/2$ with $T$ being the period of the numerator, $1-\cos{[(\varepsilon_+(p_{\rm max})-\varepsilon_-(p_{\rm max}))T]}=0$.}
\bibitem{footnote:multiband} This criterion only applies to the multi-band case. In the single-band case, other ways to distinguish the two probe pulses than energy resolution have to be employed. This could, for instance, be polarized probe pulses in combination with selection rules.
\end{thebibliography}
\end{document}